\documentclass[onecolumn,preprintnumbers,amsmath,amssymb,aps]{revtex4}
 \normalsize

\usepackage[dvips]{color}
\usepackage{graphicx,graphics}
\usepackage{dcolumn}
\usepackage{bm}
\usepackage{amssymb}
\usepackage{amsmath}
\newcommand{\be}{\begin{eqnarray}}
\newcommand{\ee}{\end{eqnarray}}

\def\bea{\begin{eqnarray}}
\def\eea{\end{eqnarray}}
\newcommand{\bi}{\bibitem}
\def\lsim{\raise0.3ex\hbox{$\;<$\kern-0.75em\raise-1.1ex\hbox{$\sim\;$}}}
\def\gsim{\raise0.3ex\hbox{$\;>$\kern-0.75em\raise-1.1ex\hbox{$\sim\;$}}}

\def\ie{{\it i.e.}}

\begin{document}
\title{Muon Anomalous Magnetic Moment and \fontsize{10}{10}{$\mu \to e \gamma$} in \fontsize{10}{10}{$B-L$} Model with Inverse Seesaw}

\author{W. Abdallah$^{1,2}$, A. Awad$^{1,3}$, S. Khalil$^{1,4,5}$, and H. Okada$^{6}$}
\affiliation{$^1$ Centre for Theoretical Physics, Zewail City of Science and Technology, Sheikh Zayed, 12588, Giza, Egypt.\\
$^2$ Department of Mathematics,  Faculty of Science, Cairo
University, Giza, Egypt.\\
$^3$ Department of Physics, Faculty of Science, Ain Shams
University, Cairo, Egypt.\\
$^4$ Department of Mathematics, Faculty of Science, Ain Shams
University, Cairo, Egypt.\\
$^5$ Centre for Theoretical Physics, British University in Egypt,
El Sherouk City, 11837, Egypt.\\
$^6$ School of Physics, KIAS, Seoul 130-722, Korea.} 

\begin{abstract}
We study the anomalous magnetic moment of the muon, $a_\mu$, and
lepton flavor violating decay $\mu \to e \gamma$ in TeV scale
$B-L$ extension of the Standard Model (SM) with inverse seesaw
mechanism. We show that the $B-L$ contributions to $a_\mu$ are
severely constrained, therefore the SM contribution remains
intact.  We also emphasize that the current experimental limit of
$BR(\mu \to e \gamma)$ can be satisfied for a wide range of
parameter space and it can be within the reach of MEG experiment.
\end{abstract}

\maketitle
\section{Introduction}

The anomalous magnetic moment of the muon has been measured at
 Brookhaven National Laboratory to a precision of $0.54$ parts per million. The current
average of the experimental results is given by \cite{bennett}
\be%
a^{\rm exp}_{\mu}=11 659 208.0(6.3)\times 10^{-10},%
\ee%
which is different from the Standard Model (SM) prediction by $3.3
\sigma$ to $3.6\sigma$ \cite{discrepancy1,discrepancy2}
\be %
\Delta a_{\mu}=a^{\rm exp}_{\mu}-a^{\rm SM}_{\mu}=(28.3 \pm
8.7\ {\rm to}\ 28.7 \pm
8.0)\times 10^{-10}.
\ee%

This discrepancy has been established by the impressive accuracy
of recent theoretical and experimental results. Therefore, it is
tempting to consider the above result as a strong signature for physics
beyond the SM. It is important to note that the SM estimation for
$a_\mu$ depends on the low-energy hadronic vacuum polarization,
which is the main source of the uncertainty. The above result is
obtained when  hadronic vacuum polarization is determined directly
from the annihilation of $e^+ e^-$ to hadrons. However, if
hadronic $\tau$ decays are included, substantially larger value
for $a^{\rm had}_{\mu}$ is derived that reduces the
discrepancy to about $2.4\, \sigma$ only \cite{discrepancy2}.

In addition, non-vanishing neutrino masses confirmed by neutrino
oscillation experiments \cite{osc-ex} are one of the firm
observational evidences for an extension of the SM. The simplest
way to account for small neutrino masses is to introduce
right-handed neutrinos into the SM, which are Majorana-type
particles with very heavy masses. In this case, type I seesaw
mechanism \cite{seesaw} can be implemented and an elegant
explanation for light neutrinos is obtained. Recently, it has been
shown that TeV scale right-handed neutrinos can be naturally
implemented in $B -L$ extension of the SM \cite{Khalil:2006yi},
where three SM singlet fermions arise naturally to cancel the
$U(1)_{B-L}$ triangle anomaly. Also, the scale of $B-L$ symmetry
breaking can be related to supersymmetry breaking scale
\cite{Khalil:2007dr}, therefore, the right-handed neutrino masses
are naturally of order {\rm TeV} scale.

In order to fulfill the experimental measurements for the
light neutrino masses with TeV scale right-handed neutrino, a very
small Dirac neutrino Yukawa couplings, $Y_{\nu} < {\cal
O}(10^{-7})$ must be assumed \cite{Khalil:2006yi}. In this case,
the mixing between light and heavy neutrinos are negligible, and
hence the interactions of right-handed neutrinos with the SM
particles are very suppressed. In Ref.\cite{Khalil:2010zz}, a
modification to the TeV scale $B-L$ model is proposed to prohibit
type I seesaw and allow another scenario for generating the light neutrino masses, namely the inverse seesaw mechanism
\cite{Mohapatra:1986bd,GonzalezGarcia:1988rw}. In this scenario,
the neutrino Yukawa coupling is no longer suppressed and can be
of order one. Thus, the heavy neutrinos associated to this model
are quite accessible and lead to interesting phenomenological
implications.

In this paper we analyze the anomalous magnetic moment of the muon
in TeV scale $B-L$ extension of the SM with inverse seesaw
mechanism. We provide analytical formula for loop contributions
due to the exchange of right-handed neutrinos, $B-L$ gauge boson,
and extra Higgs. We show that right-handed neutrinos give the
dominant $B-L$ contribution to $a_\mu$.
However, the unitarity violation limits of the light neutrino mixing
matrix restrict this effect significantly.
We also consider the impact of the right-handed neutrinos on the
Lepton Flavor Violation (LFV) decays $\mu \to e \gamma$. We show that
the rate of this decay is enhanced and becomes within the reach of
present experiments.

The paper is organized as follows. In section 2 we briefly review
the TeV scale gauged $B-L$ model with inverse seesaw mechanism. We
focus on the neutrino sector and show that the unitarity violation
limits of $U_{\!M\!N\!S}$ mixing matrix constrain the mixing between
light and heavy neutrinos. In section 3 we study the anomalous
magnetic moment of the muon due to the exchange of heavy
neutrinos, $B-L$ gauge boson $Z'$ and $B-L$ extra Higgs $H'$. In
section 4 we analyze the LFV process $\mu \to e \gamma$ and the
constrained imposed by the experimental limit of $BR(\mu \to e
\gamma)$ on the heavy neutrino contributions. Section 5 is devoted
for the numerical results and possible correlation between
$a_\mu$ and $BR(\mu \to e \gamma)$. Finally we give our
conclusions in section 6.

\section{TeV scale $B-L$ with Inverse seesaw}

In this section we briefly review the TeV scale $B-L$ extension of
the SM with inverse seesaw mechanism, which has been recently
proposed in Ref.\cite{Khalil:2010zz}. This model is based on the
gauge group $SU(3)_C\times SU(2)_L\times U(1)_Y\times U(1)_{B-L}$,
where the $U(1)_{B-L}$ is spontaneously broken by a SM singlet
scalar $\chi$ with $B-L$ charge $=-1$. Also a gauge boson $Z'$ and
three SM singlet fermions $\nu_{R_i}$ with $B-L$ charge $=-1$ are
introduced for the consistency of the model. Finally, three SM
singlet fermions $S_1$ with $B-L$ charge $=-2$ and three singlet
fermions $S_2$ with $B-L$ charge $=+2$ are considered to implement
the inverse seesaw mechanism.  The $B-L$ quantum numbers of
fermions and Higgs bosons of this model are given in
Table~\ref{tab:b-l}.
\begin{table}[thbp]
\centering {\large
\begin{tabular}{||c|c|c|c|c|c|c|c|c|c|c|c||}
\hline\hline ~~Particle~~ & ~~$Q$~~ & ~~$u_R$~~& ~~$d_R$~~& ~~$L$~~ & ~~$e_{R}$~~ & ~~$\nu _{R}$ ~~& ~~$\phi$~~ & ~~$\chi$~~ & ~~$S_1$~~ & ~~$S_2$ ~~\\
\hline
$Y_{B-L}$& $1/3$  & $1/3$ & $1/3$ & $-1$ & $-1$ & $-1$ & $0$ & $-1$ & $-2$ & $+2$\\
\hline\hline
\end{tabular}%
} \caption{$B-L$ quantum numbers of fermions and Higgs particles}
\label{tab:b-l}
\end{table}

The relevant part of the Lagrangian in this model is given by%
\begin{eqnarray}%
{\cal L}_{B-L}&=&-\frac{1}{4} F'_{\mu\nu}F'^{\mu\nu} + i~ \bar{L}
D_{\mu} \gamma^{\mu} L + i~ \bar{e}_R D_{\mu} \gamma^{\mu} e_R +
i~ \bar{\nu}_R D_{\mu} \gamma^{\mu} \nu_R + i~
\bar{S}_1 D_{\mu} \gamma^{\mu} S_1 + i~ \bar{S}_2 D_{\mu} \gamma^{\mu} S_2\nonumber\\
&+& (D^{\mu} \phi)^\dag (D_{\mu} \phi) + (D^{\mu} \chi)^\dag
(D_{\mu} \chi)- V(\phi, \chi)- \Big(\lambda_e \bar{L} \phi e_R +
\lambda_{\nu} \bar{L} \tilde{\phi} \nu_R +
\lambda_{S} \bar{\nu}^c_R \chi S_2 + h.c.\Big)\nonumber\\
&-& \frac{1}{M^3}\bar{S}^c_{1} {\chi^\dag}^{4} S_{1}-\frac{1}{M^3}\bar{S}^c_{2} {\chi}^{4} S_{2},%
\label{lagrangian}
\end{eqnarray} %
where $F'_{\mu\nu} = \partial_{\mu} Z'_{\nu} - \partial_{\nu}
Z'_{\mu}$ is the field strength of the $U(1)_{B-L}$. The general expression for the covariant
derivative $D_{\mu}$ is defined as%
\be%
D_{\mu} = \partial_\mu - i g_s  T^a G_\mu^a- i g \frac{\tau^i}{2} W_\mu^i -
i g' Y B_\mu -  i g'' Y_{B-L} Z'_{\mu}, %
\ee %
where $g''$ is the $U(1)_{B-L}$ gauge coupling constant. The last
two terms in ${\cal L}_{B-L}$ are non-renormalizable terms, which
are allowed by the symmetries and relevant for generating small
mass for $S_1$ and $S_2$ at TeV, are required by inverse seesaw
mechanism. Few remarks are in order: $i)$ The $B-L$ symmetry
allows a mixing kinetic term $F_{\mu\nu}F'^{\mu\nu}$. This term
leads to a mixing between $Z$ and $Z'$. However due to the
stringent constraint from LEP II on $Z-Z'$ mixing, one may neglect
this term. In our analysis we assume a minimal model of $B-L$
extension of the SM. $ii)$ In order to avoid other possible
non-renormalizable term that may spoil the inverse seesaw
mechanism that we adopt, a discrete symmetry like $Z_4$ is
imposed. $iii)$ In order to avoid a large mass term $m_s S_1 S_2$
in the above Lagrangian, one assumes that the SM particles,
$\nu_R$, $\chi$, and $S_2$ are even under a $Z_2$-symmetry, while
$S_1$ is an odd particle. Finally, $V(\phi, \chi)$ is the most
general Higgs potential invariant under these symmetries and it is
given by \cite{Khalil:2006yi}%
\begin{equation}%
V(\phi,\chi)=m_1^2 \phi^\dagger \phi+m_2^2 \chi^\dagger
\chi+\lambda_1 (\phi^\dagger\phi)^2+\lambda_2
(\chi^\dagger\chi)^2
+\lambda_3
(\chi^\dagger\chi)(\phi^\dagger\phi),%
\end{equation}%
where $\lambda_3 > - 2 \sqrt{\lambda_1\lambda_2}$ and $\lambda_1,
\lambda_2 \geq 0$, so that the potential is bounded from below.\\

The non-vanishing vacuum expectation value (vev) of $\chi$:
$|\langle\chi\rangle|= v'/\sqrt 2$ is assumed to be of order TeV,
which is consistent with the result of radiative $B-L$ symmetry
breaking found in gauged $B-L$ model with supersymmetry
\cite{Khalil:2007dr}. The vev of the Higgs field $\phi$:
$|\langle\phi^0\rangle|= v/\sqrt 2$ breaks the electroweak (EW)
symmetry, \ie, $v =246$ GeV. After the $B-L$ gauge symmetry
breaking, the gauge field $Z'$ acquires the following mass:
\begin{equation}
M^2_{Z'}=g''^2 v^{\prime 2}.
\end{equation}%
The experimental search for $Z'$ LEP II \cite{m.carena}
leads to
\begin{equation}
M_{Z'}/g''>6\ {\rm TeV}.
\end{equation}%

Also, after the $B-L$ and the EW symmetry breaking, the
neutrino Yukawa interaction terms lead to the following mass
terms:%
\be%
{\cal L}_m^{\nu} = m_D \bar{\nu}_L \nu_R + M_N \bar{\nu}^c_R S_2 + h.c.,%
\ee%
where $m_D=\frac{1}{\sqrt{2}}\lambda_\nu v$ and $ M_N =
\frac{1}{\sqrt 2}\lambda_{S} v' $. In addition the second
non-renormalizable term in Eq.(\ref{lagrangian}) induces a
Majorana mass for $S_2$ fermion.
Hence, the Lagrangian of neutrino masses, in the flavor basis, is given by %
\be%
{\cal L}_m^{\nu} =\mu_s \bar{S}^c_2 S_2 +(m_D \bar{\nu}_L \nu_R + M_N \bar{\nu}^c_R S_2 +h.c.) ,%
\ee%
where $\mu_s=\frac{v'^4}{4 M^3}\sim10^{-9}$ GeV, hence $M$ is of
order intermediate scale $10^7$ GeV and the  flavor indices are
omitted for simplicity. Therefore, the neutrino mass matrix can be
written as ${\cal M}_{\nu} \bar{\psi}^c \psi$
with $\psi=(\nu_L^c ,\nu_R, S_2)$ and ${\cal M}_{\nu}$ is given by %
\be {\cal M}_{\nu}=
\left(%
\begin{array}{ccc}
  0 & m_D & 0\\
  m^T_D & 0 & M_N \\
  0 & M^T_N & \mu_s\\
\end{array}%
\right). %
\ee%
The diagonalization of this mass matrix
leads to the following light and heavy neutrino masses, respectively: %
\begin{eqnarray}%
m_{\nu_l} &=& m_D M_N^{-1} \mu_s (M_N^T)^{-1} m_D^T,\label{mnul}\\
m_{\nu_H}^2 &=& m_{\nu_{H'}}^2 = M_N^2 + m_D^2. %
\end{eqnarray} %

It is now clear that the light neutrino masses can be of order eV,
with a TeV scale $M_N$ if $\mu_s \ll M_N$. Therefore, the Yukawa
coupling $\lambda_\nu$ is no longer suppressed and can be of order one. Such large coupling is a crucial
for testing this type of models and probing the heavy neutrino physics at LHC, as shown in Ref.\cite{kh}. The light neutrino mass matrix
in Eq. (\ref{mnul}) must be diagonalized by the physical neutrino
mixing matrix $U_{\!M\!N\!S}$ \cite{mns}, {\it i.e.}, %
\be%
U_{\!M\!N\!S}^T m_{\nu_l} U_{\!M\!N\!S} = m_{\nu_l}^{\rm diag} \equiv
{\rm diag}\{m_{\nu_e}, m_{\nu_\mu}, m_{\nu_\tau}\}.%
\ee
Thus, one can easily show that the Dirac neutrino mass
matrix can be defined as :%
\be %
m_D=U_{\!M\!N\!S}\, \sqrt{m_{\nu_l}^{\rm diag}}\, R\, \sqrt{\mu^{-1}_s}\, M_N, %
\label{mD}
\ee %
where $R$ is an arbitrary orthogonal matrix. It is clear that this
expression is a generalization to the expression of $m_D$ in type
I seesaw, which is given by $ m_D = U_{\!M\!N\!S}\, \sqrt{m_{\nu_l}^{\rm
diag}}\, R\, \sqrt{M_N}$ \cite{Casas:2001sr}. Accordingly, the
matrix $V$ that diagonalizes the $9\times 9$ neutrino mass matrix
${\cal M}_\nu$, {\it i.e.}, $V^T {\cal M}_\nu V = {\cal
M}_\nu^{\rm diag}$, is given by \cite{Dev:2009aw}%
\be V=
\left(%
\begin{array}{cc}
  V_{3\times 3} & V_{3\times6}\\
  V_{6\times 3} & V_{6\times6}  \\
\end{array}%
\right),%
\ee%
where the matrix $V_{3\times3}$ is given by %
\be%
V_{3\times3} \simeq \left(1-\frac{1}{2} F F^T \right)
U_{\!M\!N\!S} \, . %
\ee%

It is clear that in general $V_{3\times 3}$ is not unitary matrix
and the unitarity violation, \ie, the deviation from the standard
$U_{\!M\!N\!S}$, is measured by the size of $\frac{1}{2} F F^T$. The
matrix $V_{3\times6}$ is defined as %
\be%
V_{3\times6}=\left({\bf 0}_{3\times3},F \right) V_{6\times6},\quad
~~~  F = m_D M^{-1}_N. %
\label{V36}
\ee %
Finally, $V_{6\times 6}$ is the matrix that diagonalize the
$\{\nu_R, S_2\}$ mass matrix. Note that due to the Higgs mixing
term in the potential $V(\phi,\chi)$, the physical Higgs scalars
$(H,H')$ are given as a linear combination of $\phi$ and $\chi$,
with the following masses \cite{Khalil:2006yi}:
\be%
m^2_{H,H'} = \lambda_1 v^2 + \lambda_2 v'^2 \mp
\sqrt{(\lambda_1 v^2 - \lambda_2 v'^2)^2 +\lambda_3^2 v^2 v'^2}.%
\ee%
A detailed analysis for the phenomenology of the Higgs bosons of
this model at the LHC has been considered in Ref.
\cite{Emam:2007dy,Basso}.
\section{$B-L$ contributions to the Muon Anomalous Magnetic Moment}

\begin{figure}[t]
\includegraphics[width=4.5cm,height=3cm]{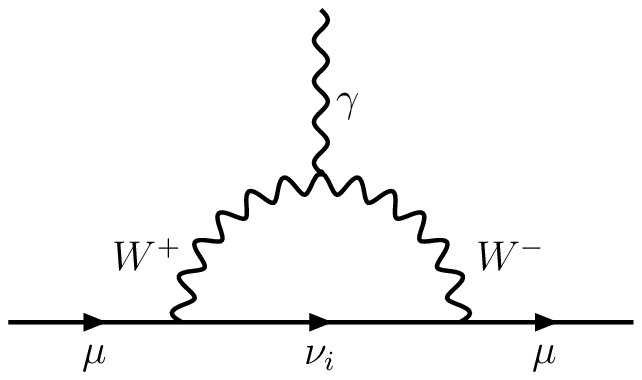}~~~~~
\includegraphics[width=4.5cm,height=3cm]{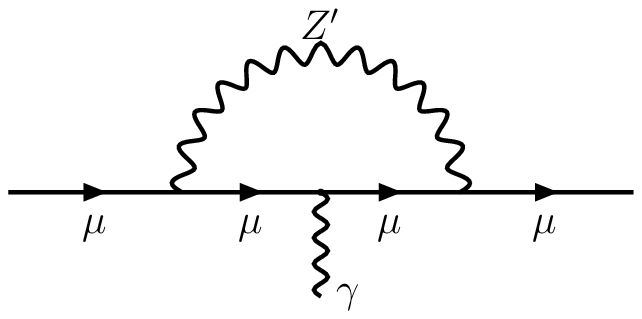}~~~~~
\includegraphics[width=4.5cm,height=3cm]{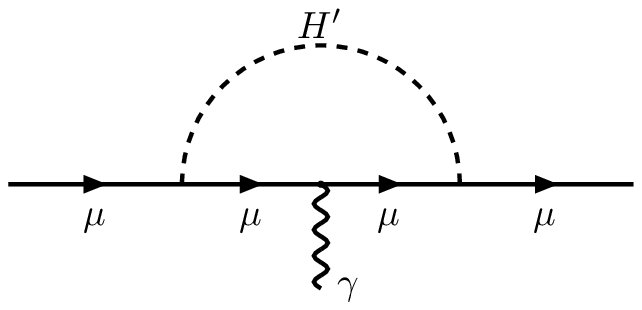}
\caption{The new contributions of the muon anomalous magnetic
moment in $B-L$ extension of the SM.} \label{g-2}
\end{figure}

In this section we analyze new contributions to the muon
anomalous magnetic moment due to the extra particles of the $B-L$ TeV scale model.
From the effective Lagrangian of leptonic sector, one finds the following interactions%
\bea%
{\cal L}= \frac{g}{\sqrt2}(V^*_{\mu i}
{\bar\nu_i} \gamma^\alpha W_{\alpha}^{+}P_L \mu + V_{\mu i}{\bar
\mu}\gamma^\alpha W_{\alpha}^{-}P_L\nu_i) + g'' \bar{\mu} \gamma^\alpha Z'_\alpha \mu
+ \lambda_\mu \sin\theta \bar{\mu} H' \mu ,
\eea %
where $V$ is $9 \times 9$ extended MNS matrix, as
discussed above, $\lambda_{\mu}$ is the Yukawa coupling of the
muon, and $\theta$ is the mixing angle between the SM-like Higgs
and extra Higgs \cite{Emam:2007dy}. Thus, one can easily observe
that the new contributions to the anomalous magnetic moment of the
muon are generated by one loop diagrams involving the exchange of
$W$ gauge boson and heavy neutrino, or $Z'$ boson and $\mu$
exchange, or $H'$ neutral scalar boson and $\mu$, as shown in Fig.
\ref{g-2}.  Therefore, one can define $a_{\mu}^{B-L}$ as
\be %
a_{\mu}^{B-L}= a^{\nu}_{\mu}+ a^{Z'}_{\mu}+ a^{H'}_{\mu}. \ee

The calculation of the first diagram in Fig. \ref{g-2} leads to   %
\be%
a^{\nu}_{\mu}= \frac{G_F}{\sqrt2}\frac{m^2_{\mu}}{8\pi^2}\sum^{9}_{i=1}V^*_{\mu i}V_{\mu i} f(r_{\nu_i}),
\ee%
where $r_{\nu_i}= (m_{\nu_i}/M_W)^2$ and $f(r)$ is given by%
\be%
f(r)= \frac{10-43r+78r^2-49r^3+4r^4+18r^3\ln (r)}{3(1-r)^4}.
\ee%
In this calculation, we assume that $(m_{\mu}/M_W)^2\simeq0$. %
For $r \simeq 0$ one finds that $f(0)=10/3$, while if $r \gg
1$ then $f(r) \to 4/3$.  Thus, in the SM this contribution implies %
\be%
(a_\mu^\nu)^{\rm SM} = \frac{G_F}{\sqrt2}\frac{m^2_{\mu}}{8\pi^2}
\times \frac{10}{3} = 3.89 \times 10^{-9},%
\ee%
where the mixing matrix $V$ is given by the unitary $U_{\!M\!N\!S}$
mixing matrix, {\it i.e.}, $\sum_{i=1}^3 \vert U_{\mu i}\vert^2
=1$. In our model with TeV scale $B-L$, the $9\times 9$ mixing
matrix $V$ is unitary, however the $3 \times 3$ mixing matrix of
light neutrino is no longer unitary. In our analysis, we
constrain ourselves with the following non-unitary limits for light neutrino mixing matrix \cite{Antusch:2006vwa}%
\begin{equation}\label{Nm}
\vert N N^\dagger\vert \approx \left(%
\begin{array}{ccc}
0.994 \pm 0.005 & ~~<7.0\times10^{-5} & ~~<1.6\times10^{-2}\\
<7.0\times10^{-5} & ~~0.995 \pm 0.005 & ~~<1.0\times10^{-2}\\
<1.6\times10^{-2} & ~~<1.0\times10^{-2} & ~~0.995 \pm 0.005
\end{array}%
\right).%
\end{equation}
In this case, one finds that  $0.99 \leq \sum_{i=1}^3 \vert
V_{\mu i} \vert^2 \leq 1$. Hence the SM-like contribution is
slightly reduced to $3.851 \times 10^{-9}$. Since $4/3\leq f(r)\leq 10/3$, one can
easily see that %
\begin{eqnarray*}
\frac{10}{3} \geq \sum_{i=1}^9 \vert V_{\mu i}\vert^2 f(r_i)
&\geq & \frac{10}{3} \sum_{i=1}^3 \vert V_{\mu i}\vert^2 +
\frac{4}{3}\left(1 - \sum_{i=1}^3 \vert V_{\mu i}\vert^2 \right)
\nonumber\\
& \geq & \frac{6}{3} \sum_{i=1}^3 \vert V_{\mu i}\vert^2 +
\frac{4}{3}. %
\end{eqnarray*}
Thus, the ratio $R_\mu^\nu = a_\mu^\nu/(a_\mu^\nu)^{\rm SM}$ lies within the tiny range:%
\be%
0.994 \leq R_\mu^\nu \leq 1, %
\ee%
which means that within TeV scale $B-L$ extension of the SM, the
discrepancy between the theoretical prediction of the anomalous
magnetic moment of the muon and its experimental measurement
remains $2.4 \sigma$ as in the SM and another source of new
physics is required to account for this difference.


Next, we consider the contribution of $Z'$ to $a_{\mu}$. From the second diagram in Fig. \ref{g-2}, one finds the following result:
\be%
a^{Z'}_{\mu}= \frac{g''^2}{4\pi^2}\frac{m^2_{\mu}}{M^2_{Z'}} g(r_{m_\mu}),%
\ee%
where we assume that $r_{m_\mu} \equiv
(m_\mu/M_{Z'})^2 \simeq 0$, and $g(r_{m_\mu})$ is given by%
\be%
g(r_{m_\mu})= \int^1_0dz\frac{z^2(1-z)}{1-z+r_{m_\mu}z^2}\xrightarrow{r_{m_\mu}\approx 0}\frac{1}{3}.
\ee%
Hence one finds that%
\bea %
a^{Z'}_{\mu}\approx \frac{m^2_{\mu}}{4\pi^2}
\left(\frac{g''}{M_{Z'}}\right)^2\frac13 <
 \frac{m^2_{\mu}}{12\pi^2}
\left(\frac{1}{6000\ {\rm GeV}}\right)^2 \simeq 2.34\times 10^{-12}.
\eea
This contribution is quite small and one can neglect the effect of the $Z'$ diagram.
Finally, we consider the diagram of extra Higgs. We find that the corresponding contribution to $a_\mu$ is given by%
\be%
a^{H'}_{\mu}= \frac{|\lambda_\mu \sin\theta|^2}{32\pi^2}\frac{2m^2_{\mu}}{m^2_{H'}} h(r'_{m_\mu}),%
\ee
where $r'_{m_\mu} = (m_\mu/m_{H'})^2 \simeq 0$ is assumed and $\lambda_\mu = m_\mu/v \simeq {\cal O}(10^{-4})$.
The loop function $h(r'_{m_\mu})$ is given by%
\be%
h(r'_{m_\mu})= \frac{2+3r'_{m_\mu}-6r'^2_{m_\mu}+r'^3_{m_\mu}+6r'_{m_\mu}\ln (r'_{m_\mu})}{3(1-r'_{m_\mu})^4}.
\ee%
Hence one can estimate this contribution, for $m_{H'} \simeq {\cal O}(100)$ GeV and $\sin\theta=1/\sqrt2$, as %
\be%
a^{H'}_{\mu} \simeq  10^{-16}, %
\ee%
which is also quite negligible. Therefore, one concludes that the
$B-L$ contributions to $a_\mu$ can not account for the reported
discrepancy between the theoretical and experimental expectations.
It is worth noting that $B-L$ contribution to $a_{\mu}$, obtained
from the loop diagram mediated by heavy neutrinos can be of order
the SM contribution and has a significant effect if the mixing
between light and heavy neutrinos is sizable. However this mixing
is strongly constrained by several leptonic processes
\cite{Antusch:2006vwa}.

\section{\fontsize{10}{10}{$\mu$}$\to e\gamma$ in TeV scale \fontsize{10}{10}{$B-L$} with Inverse seesaw}

We now consider the LFV decay $\mu \to e \gamma$ in the TeV scale
$B-L$ model with inverse seesaw mechanism. Many experiments have
been designed to search for LFV processes, in particular $\mu \to
e \gamma$. The current upper
limit is given by the MEG experiment \cite{meg}%
\be%
BR(\mu \to e \gamma) < 2.4 \times 10^{-12}.
\label{mubound}
\ee%
New experiments are expected to improve this limit by three order
of magnitudes. It is important to note that the SM result for the
branching ratio of $\mu \to e \gamma$, with neutrino masses as in Eqs.(\ref{eq1},\ref{eq2}), is given by%
\be%
BR(\mu \to e \gamma)^{\rm SM} \simeq  10^{-55}.%
\ee%
Thus, the observation of $\mu \to e \gamma$ decay will be a
clear signal for physics beyond the SM.
\begin{figure}[t]
\includegraphics[width=4.5cm,height=3cm]{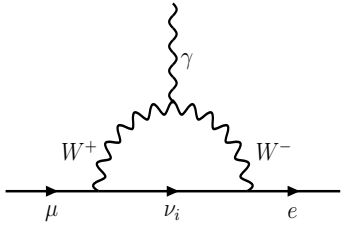}
\caption{$\mu\rightarrow e \gamma$ in $B-L$ extension of the SM.}
\label{muegamma}
\end{figure}
We perform the calculation of $\mu \to e \gamma$ due to the
exchange of light and heavy neutrinos and $W$ gauge boson inside
the loop, as shown in Fig. \ref{muegamma}. The amplitude of $\mu
\to e
\gamma$, in the limit of $m_e \to 0$, can be written as%
\be %
A(\mu\rightarrow e \gamma) \simeq
\frac{m_{\mu}G_F}{32\sqrt2\pi^2}\sum_{i=1}^9{V^*_{\mu i}V_{e i} f(r_i)
\times \bar u(p)[2e(p'\cdot \epsilon)] u(p')}. \ee
Let us define the coefficient of the amplitude $A$ as %
\bea%
a=\frac{em_{\mu}G_F}{32\sqrt2\pi^2}\sum_{i=1}^9{V^*_{\mu i}V_{e
i}f(r_i)}.%
\eea%
Therefore, the decay rate is given by %
\bea%
\Gamma(\mu \rightarrow e \gamma)=\frac{m^3_{\mu}}{8\pi}|a|^2.%
\eea %
Using the dominant decay mode of $\Gamma(\mu\rightarrow e \nu
\bar\nu)\simeq
m^5_{\mu}G^2_F/(192\pi^3)$, the branching ratio is given by %
\begin{eqnarray}
BR(\mu\rightarrow e \gamma) = \frac{\Gamma(\mu\rightarrow e
\gamma)}{\Gamma(\mu\rightarrow e \nu \bar\nu)}
=\frac{m^3_{\mu}|a|^2}{8\pi} \frac{192\pi^3}{m^5_{\mu}G^2_F}=
\frac{3\alpha}{64 \pi} \left\vert \sum^{9}_{i=1}  V^*_{\mu i}V_{e
i} f(r_i) \right\vert^2,
\end{eqnarray}
where $\alpha=e^2/4\pi\simeq 1/137$. From the experiment upper
bound in Eq.(\ref{mubound}), one finds the following constraint on
the light and heavy neutrino mixing:%
\be\label{bound-mu}
\left|\sum^{9}_{i=1} V^*_{\mu i}V_{e i} f(r_i)\right| < 0.000149. %
\ee%

In case of extremely heavy right-handed neutrinos, the lepton
mixing matrix $V_{3\times 3}$ is almost unitary. Therefore, the
contribution of light neutrinos, which corresponds to $i=1,2,3$,
is almost zero. In addition, the contribution of heavy neutrinos
is quite suppressed due to the small mixing between light and
heavy neutrinos in this case ($V_{\mu i} \sim V_{e i} \sim m_D/M_N
\sim {\cal O}(10^{-9})$). Hence the above constraint is satisfied
and $BR(\mu \to e \gamma) \ll 10^{-12}$ is obtained.

However, within  TeV scale inverse seesaw the lepton mixing matrix
is non-unitary. Also the mixing between heavy and light neutrinos
are not small, since $m_D/M_N \sim {\cal O}(0.1)$. Therefore, the
bound in
Eq.(\ref{bound-mu}) can be written as %
\be%
\left|\frac{10}{3} \sum_{i=1}^3 V^*_{\mu i} V_{ei} + \sum_{j=4}^9
V^*_{\mu j} V_{ej} f(r_j)\right| < 0.000149, %
\ee%
where $r_j =(m_{\nu_{H_j}}/M_W)^2$ and $V_{e(\mu)j}$, as defined
in Eq.(\ref{V36}), is given by%
\begin{eqnarray}%
V_{e(\mu) j} = \left[\left(0, m_D M^{-1}_N\right) V_{6\times6}
\right]_{1(2),j-3} = \left[\left(0,U_{\!M\!N\!S}\sqrt{m_{\nu_l}^{\rm diag}
} R \sqrt{\mu^{-1}_s}\right) V_{6\times6}\right]_{1(2),j-3}.
\label{Vmui}
\end{eqnarray}
Thus, for $r_j \gg 1$, {\it
i.e.} $f(r_j) =4/3$, one finds %
\be%
\left|\sum_{i=1}^3 V^*_{\mu i} V_{ei}\right|< 0.0000636, %
\ee which implies that $(F F^T)_{21,12} \lsim 10^{-4}$. This bound
can be easily satisfied, due to the constraints imposed on the
off-diagonal elements of the non-unitary $U_{\!M\!N\!S}$ mixing matrix.

\section{Numerical results}

In our model of $B-L$ extension of the SM with inverse seesaw
mechanism, the relevant input parameters involved in the computation
of the anomalous magnetic moment of muon are the following:%
\begin{enumerate}
\item Three right-handed neutrino masses.
\item Three $\mu_s$ mass parameters.
\item Three angles of the orthogonal matrix $R$.
\end{enumerate}

In fact, the form of the Dirac neutrino mass matrix $m_D$, given in Eq.(\ref{mD}), guarantees that we obtain the correct
light neutrino masses and mixing matrix $U_{\!M\!N\!S}$.

The solar and atmospheric neutrino oscillation experiments provide the following results for the neutrino mass-squared differences with best-fit values within 1$\sigma$ errors \cite{Schwetz:2011qt}:%
\begin{eqnarray}
\Delta m^2_{21} &=& (7.64^{+0.19}_{-0.18})\times 10^{-5}~{\rm eV}^2,\label{eq1} \\
\vert \Delta m^2_{31}\vert &=& (2.45\pm0.09) \times 10^{-3}~{\rm eV}^2.\label{eq2}
\end{eqnarray}%
Therefore, one finds %
\bea%
m_{\nu_{l_2}}&=& \sqrt{7.64\times 10^{-5} + m_{\nu_{l_1}}^2},\\
m_{\nu_{l_3}}&=& \sqrt{\vert 2.45\times 10^{-3}+
m_{\nu_{l_1}}^2 \vert},%
\eea%
with arbitrary $m_{\nu_{l_1}}$. If  $m_{\nu_{l_1}}^2 \ll 7.64 \times
10^{-5}{\rm eV}^2$, the ansatz of hierarchal the light neutrino masses is obtained.

In this case one gets the following the light neutrino masses:%
\be%
m_{\nu_{l_1}} \lsim 10^{-5} ~ {\rm eV},~~~~~~  m_{\nu_{l_2}} = 0.008 ~ {\rm eV},~~~~~~
m_{\nu_{l_3}} = 0.05 ~{\rm eV}.%
\ee%

In our analysis, we adopt these values for the light neutrino masses
and also assume that the neutrino mixing matrix is given by
Eq.(\ref{Nm}). From Eq.(\ref{Vmui}), one notices that the mixing
element $V_{\mu i}$, which plays a crucial role in the result of
$a_{\mu}$, can be enhanced if: $(i)$ $\mu_{s_i} \lsim
m_{\nu_{l_i}}$, $(ii)$ the orthogonal matrix $R$ is maximally
mixing. For example, if $\mu_{s_3}=2.7\times10^{-9}\;{\rm GeV}$, $M_{N_1}=900\,{\rm GeV}$,
$M_{N_2}=1500\,{\rm GeV}$, $M_{N_3}=1900\,{\rm GeV}$, and the other parameters are fixed as in Fig.
\ref{Br1}, then one finds the following
extended MNS mixing matrix: %
\begin{figure}[t]
\includegraphics[width=10cm,height=6.5cm]{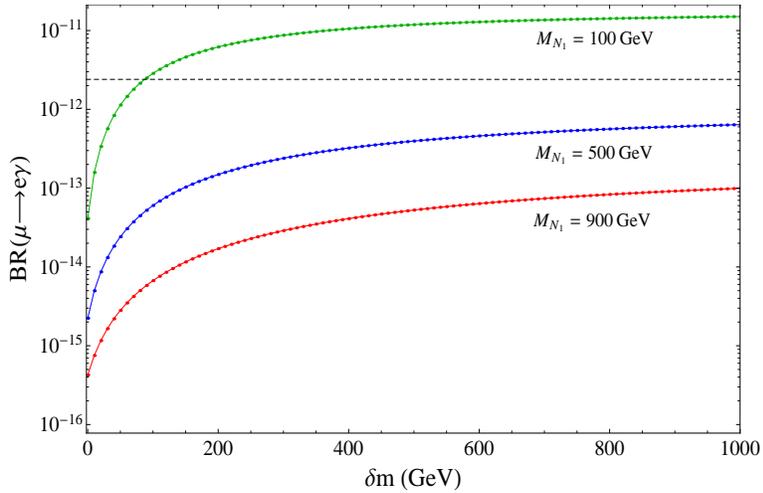}
\caption{$BR(\mu\rightarrow e \gamma)$ versus $\delta
m=M_{N_2}-M_{N_1}$, for $M_{N_1}=100, 500, 900\;{\rm GeV}$ from up
to down, respectively. The horizontal dashed line refers to the
MEG experiment limit of $BR(\mu \to e \gamma)$. The other
parameters are fixed as follows: $M_{N_3}=2000\;{\rm GeV},\,
\mu_{s_{1}}=10^{-10}\;{\rm GeV},\, \mu_{s_{2}}=10^{-8}\;{\rm
GeV},\, \mu_{s_{3}}=2.62\times10^{-9}\;{\rm GeV},\,
m_{\nu_{l_{1}}}=10^{-13}\;{\rm GeV},\,
m_{\nu_{l_{2}}}=8.5\times10^{-12}\;{\rm GeV},\,
m_{\nu_{l_{3}}}=5.05\times10^{-11}\;{\rm GeV}$.} \label{Br1}
\end{figure}
\begin{eqnarray}%
V\simeq{\small\left(
\begin{array}{ccccccccc}
 0.806 & -0.591 & 0.001 & -0.008 & -0.008 & -0.012 & -0.012 & 0.0001 & -0.0001 \\
 -0.418 & -0.569 & 0.701 & 0.020 & 0.020 & -0.012 & -0.012 & 0.068 & -0.068 \\
 0.417 & 0.570 & 0.701 & -0.020 & -0.020 & 0.012 & 0.020 & 0.068 & -0.068 \\
 0 & 0 & 0 & 0.707 & -0.707 & -0.001 & 0.001 & 0 & 0 \\
 0 & 0 & 0 & 0.001 & -0.001 & 0.707 & -0.707 & 0 & 0 \\
 0 & 0 & 0 & 0 & 0 & 0 & 0 & -0.707 & -0.707 \\
 -0.032 & -0.026 & 0 & -0.707 & -0.707 & 0 & 0 & 0 & 0 \\
 0 & 0.029 & 0 & -0.001 & -0.001 & -0.707 & -0.707 & 0 & 0 \\
 0 & 0 & -0.136 & 0 & 0 & 0 & 0 & 0.701 & -0.701
\end{array}
\right)}.
\end{eqnarray}%

From this example, one notices that the elements $V_{\mu i}$,
$i=4,..,9$ are of order ${\cal O}(0.01)$ which induce the SM-like
contribution $3.846\times10^{-9}$ and total contribution of order
$3.863 \times 10^{-9}$. Thus, the ratio $R_\mu^\nu=
a_\mu^\nu/(a_\mu^\nu)^{\rm SM}$ is given by  $R_\mu^\nu=0.994$.

\begin{figure}[t]
\includegraphics[width=10cm,height=6.5cm]{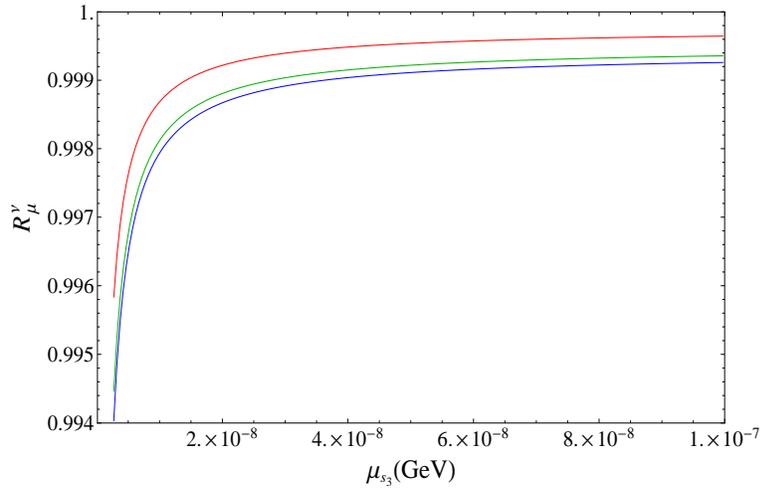}
\caption{The ratio between the SM and $B-L$ contributions to the
anomalous magnetic moment of muon, $R_\mu^\nu$, as a function of
$\mu_{s_{3}}$, for $M_{N_i}=(120,\,125,\,300)\,{\rm
GeV},$ $ (400,\,465,\,800)\,{\rm GeV}$ and $(900,\,1500,\,
1900)\,{\rm GeV}$ from up to down, respectively. The other
parameters are fixed as in Fig. \ref{Br1}.} \label{amu}
\end{figure}

In the Fig. \ref{Br1}, we present the $BR(\mu\to e \gamma)$ versus
$\delta m=M_{N_2}-M_{N_1}$ for $M_{N_1}=100, 500, 900\; {\rm GeV}$
from up to down, respectively. Here we assume that
$M_{N_3}=2000\;{\rm GeV},\, \mu_{s_{1}}=10^{-10}\;{\rm GeV},\,
\mu_{s_{2}}=10^{-8}\;{\rm GeV},\,
\mu_{s_{3}}=2.62\times10^{-9}\;{\rm GeV},\,
m_{\nu_{l_{1}}}=10^{-13}\; {\rm GeV},\,
m_{\nu_{l_{2}}}=8.5\times10^{-12}\;{\rm GeV},\,
m_{\nu_{l_{3}}}=5.05\times10^{-11}\;{\rm GeV}$. Note that the
$BR(\mu\to e \gamma)$ is not sensitive to the value of $M_{N_3}$.
For other LFV processes like $\tau\to e \gamma$ and $\tau\to \mu
\gamma$, the present experimental limits of their branching ratios
are given by \cite{BABAR}
\[BR(\tau\to e \gamma)<3.3\times10^{-8},\;\;\;\;\;\;\;BR(\tau\to \mu \gamma)<4.4\times10^{-8}.\]
One can easily show that in our models these experimental bounds
can be translated into the following constraints:
\[\left|\sum_{i=1}^3 V^*_{\tau i} V_{ei}\right|< 7.5\times10^{-3},\;\;\;\;\;\left|\sum_{i=1}^3 V^*_{\tau i} V_{\mu i}\right|<8.6\times10^{-3}.\]
This implies that $(F F^T)_{31,13} \lsim 1.5\times10^{-2}$ and $(F
F^T)_{23,32} \lsim 1.7\times10^{-2}$. As mentioned above, $F=m_D
M^{-1}_N\sim{\cal O}(0.1)$, therefore these constraints are
naturally satisfied in our model, for $m_D\sim{\cal O}(100){\rm
GeV}$ and $M_N\sim{\cal O}({\rm TeV})$ and no constraint will be
imposed.

In Fig. \ref{amu}, we plot $R_\mu^\nu$ as a function of
$\mu_{s_{3}}$, for $M_{N_i}=(120,\,125,\,300)\,{\rm
GeV},$ $ (400,\,465,\,800)\,{\rm GeV}$ and $(900,\,1500,\,
1900)\,{\rm GeV}$ from up to down, respectively. The other
parameters are fixed as in Fig. \ref{Br1}. In addition we vary the
angles of the orthogonal matrix $R$ from $0$ to $\pi$. As can be
seen from this figure, there is no significant difference between
the SM expectation and the total result of $g-2$ in TeV scale
$B-L$ extension of the SM with inverse seesaw.

\section{Conclusions}

In this paper we have computed the anomalous magnetic moment of
the muon in TeV scale $B-L$ extension of the SM with inverse
seesaw mechanism. The one loop contributions due to the exchange
of right-handed neutrinos, $B-L$ gauge boson, and extra Higgs have
been analyzed. We showed that right-handed neutrinos may give a
significant contribution to $a_\mu$, however it turns out that it
is quite restricted by sever constraints from leptonic processes.
Therefore, the SM contribution in $B-L$ extension of the SM with
inverse seesaw mechanism remains intact. Thus the discrepancy
between the theoretical prediction of $a_\mu$ and its experimental
measurement requires a different source of new physics, like
supersymmetric models with minimal flavor violation, which usually
respect the LFV constraints. We also studied the impact of the
right-handed neutrinos on the LFV decay $\mu \to e \gamma$. We
have shown that the rate of this decay is enhanced and is
reachable by the MEG experiment.

\section*{Acknowledgments}

This work is partially supported by the Science and Technology
Development Fund (STDF) project ID 1855 and the ICTP project ID
30. W.A. would like to thank A. Elsayed for fruitful discussions.



\end{document}